\documentclass[floatfix,nofootinbib,aps,prd,preprintnumbers,showpacs,showkeys]{revtex4-1}

\usepackage{amssymb}
\usepackage{amsmath}
\usepackage{color}
\usepackage{graphicx}
\usepackage{dcolumn}
\usepackage{bm}
\usepackage{epsfig}
\usepackage{color}

\def\figdir{./}

\newcommand\Eq[1]{Eq.~\ref{eq:#1}}
\newcommand\Fig[1]{Fig.~\ref{fig:#1}}

\newcommand\Tab[1]{Table~\ref{tab:#1}}

\newcommand\calO{\mathcal O}

\newcommand\calP{\mathcal P}

\begin{document}
	
\preprint{MIT-CTP/4981}

\title{Scaling properties of multiscale equilibration}

\author{W. Detmold}
\affiliation{Center for Theoretical Physics, Massachusetts Institute of Technology, Cambridge, Massachusetts 02139, USA}
\author{M. G. Endres}
\affiliation{Center for Theoretical Physics, Massachusetts Institute of Technology, Cambridge, Massachusetts 02139, USA}


\date{\today}

\begin{abstract}
We investigate the lattice spacing dependence of the equilibration time for a recently proposed multiscale thermalization algorithm for Markov chain Monte Carlo simulations.
The algorithm uses a renormalization-group matched coarse lattice action and prolongation operation to rapidly thermalize decorrelated initial configurations for evolution using a corresponding target lattice action defined at a finer scale.
Focusing on nontopological long-distance observables in pure $SU(3)$ gauge theory, we provide quantitative evidence that the slow modes of the Markov process, which provide the dominant contribution to the rethermalization time, have a suppressed contribution toward the continuum limit, despite their associated timescales increasing.
Based on these numerical investigations, we conjecture that the prolongation operation used herein will produce ensembles that are indistinguishable from the target fine-action distribution for a sufficiently fine coupling at a given level of statistical precision, thereby eliminating the cost of rethermalization. 
\end{abstract}

\maketitle

\section{Introduction}

Multiscale methods have been applied successfully in a variety of ways to facilitate Markov chain Monte Carlo (MCMC) simulations in lattice quantum chromodynamics (QCD).
These applications range from Dirac operator inversion \cite{Babich:2010qb,Babich:2009pc,Frommer:2013fsa,Brannick:2014vda} to evaluation of correlation functions and other observables \cite{Luscher:2001up,Ce:2016idq,Vera:2016xpp,Ce:2016ajy}, and have resulted in significant increases in computational efficiency, and reductions in uncertainties of stochastically estimated observables.
Implementation of a multiscale algorithm for gauge field updating in lattice QCD, however, remains an open challenge despite some early progress for simpler theories \cite{Goodman:1986pv,Edwards:1990hu,Edwards:1991eg,Janke:1993et,Grabenstein:1993nh,Grabenstein:1994ze}.

Recently, we have introduced a multiscale thermalization algorithm, based on the multigrid concepts of prolongation and restriction, and inspired by renormalization-group flows, which offers the ability to rapidly initialize an ensemble of configurations for subsequent parallel evolution \cite{Endres:2015yca}.
The benefits of using this algorithm in a QCD context are severalfold.
First, it enables the generation of ensembles with well-distributed topological charge in parameter regimes where topological freezing is problematic (namely, when the lattice spacing is less than $0.05$ fm).
Although the resulting distribution is not correctly sampled according to the fine path integration measure, formally, the deviations from the correct distribution are of order the lattice spacing, provided the topology is correctly sampled on the coarse lattice.
These lattice artifacts can be investigated using multiple prolongations and in principle corrected by subsequent reweighting.
Second, with reduced thermalization overhead, evolving fine ensembles with multiple streams becomes practical and can lead to reduced communication overhead, thereby further enhancing the efficiency of gauge field generation.
Finally, since the coarse level evolution is inexpensive, it is practical to generate fully decorrelated configurations by such an algorithm, implying greater statistical power of the resulting ensemble compared to those generated conventionally at similar cost.
Multiscale thermalization had been successfully demonstrated for both quenched \cite{Endres:2015yca} and two-flavor \cite{Detmold:2016rnh} QCD.

The main focus of the present study is to quantitatively investigate the scaling properties of multiscale thermalization as a function of the lattice spacing, under the assumption that all coarse and fine lattice pairs have been properly matched using renormalization-group matching conditions.
In addition, this study examines whether multiscale thermalization techniques might be useful for circumventing the problem of critical slowing down.
In conventional MCMC simulations, the autocorrelation times associated with long-distance observables typically scale polynomially as $\tau_\textrm{int} \sim 1/a^z$ for a fixed physical volume, where $z$ is a dynamical exponent and $a$ is the lattice spacing.
For local algorithms, the updating is typically diffusive, implying a dynamical exponent $z\sim2$, although for QCD, the scaling can be far worse due to topological freezing.
For example, between $a\sim 0.1$ fm and $a \sim 0.05$ fm, the dynamical exponent is around $z\sim5$ for pure gauge theory, using both hybrid Monte Carlo (HMC) \cite{Schaefer:2010hu} and heat bath (HB) \cite{Endres:2015yca} algorithms.
Similar scaling behavior has been observed in Ref.~\cite{Schaefer:2010hu} for gauge theories with dynamical fermions, and in both cases the topological tunneling rate is expected to become exponentially suppressed farther toward the continuum limit.
In addition to multiscale thermalization, a variety of other approaches have been proposed and studied for addressing topological freezing \cite{Luscher:2011kk,Mages:2015scv,Laio:2015era,Hasenbusch:2017unr,Bonati:2017woi} and critical slowing down \cite{Cossu:2017eys,TuLattice2017} in lattice QCD \cite{Endres:2016rzj}.

Setting aside the issue of topological freezing (e.g., by considering gauge evolution within a fixed frozen topological sector, or by applying open boundary conditions \cite{Luscher:2011kk}, as it is possible in approaching zero-temperature physics), the computational cost of conventional simulations of gauge theories nonetheless grow rapidly as the continuum limit is approached due to autocorrelations in the Markov process.
By comparison, in multiscale thermalization, the relevant timescale for attaining decorrelated configurations is no longer the autocorrelation time associated with the Markov process, but rather the {\it rethermalization} time required to equilibrate a prolongated coarse ensemble of decorrelated configurations.
(Re)thermalization and autocorrelation timescales are both tied to the slow eigenmodes of the Markov transition amplitude that defines the fine scale evolution.
However, in multiscale thermalization, the starting ensemble is drawn from a prolongated coarse distribution ($\calP_\textrm{prolongated}$), which, by design of the matched coarse action, has very good overlap with the targeted fine distribution.
This implies that the initial prolongated fine distribution is nearly orthogonal to the slowest mode(s) of evolution,\footnote{A given Markov process that satisfies detailed balance has right and left eigenvalues given by $|\chi_n\rangle$ and $\langle \tilde\chi_n|$, which satisfy the orthogonality relation $\langle \tilde\chi_n|\chi_m\rangle = \delta_{nm}$.
The corresponding eigenvalues $\lambda_n$ satisfy $|\lambda_n| = e^{-1/\tau_n}$, with $\lambda_0=1$ and $|\lambda_{n+1}| \leq |\lambda_n|$.
The mode with the largest eigenvalue, $|\chi_0\rangle$, corresponds to the target distribution to be sampled.%
} $\chi_n$ ($n=1,2,\ldots$), and therefore it is possible that only the highly excited modes of evolution control the rethermalization time.
Under such conditions, it was numerically observed that the rethermalization timescale can be significantly shorter than the associated autocorrelation timescale for fine evolution~\cite{Endres:2015yca}.

If the lattice spacing dependence of the rethermalization time scales better than that for autocorrelation times in a conventional approach [e.g., $\tau_\textrm{R} = {\cal O}(1/a^{z_\textrm{R}})$ with a rethermalization exponent $z_\textrm{R}<2$, for nontopological observables], or the overlap of prolongated fine distributions onto slow modes decreases sufficiently fast with $a$, then multiscale thermalization could offer a new strategy for addressing the problem of critical slowing down.
From a theoretical standpoint, whether or not rethermalization times scale better than autocorrelation times is a nontrivial question, since it involves not only understanding the spectral properties of the transition probability matrix of the Markov process, but also its density of eigenmodes.
Although in general not much can be said theoretically about the scaling properties of either with lattice spacing (though our expectation is that the slow modes are generally diffusive for local updating schemes, and thus have quadratic scaling with inverse lattice spacing ), heuristic and perturbative arguments suggest that the overlap factors arising from multiscale thermalization will diminish with lattice spacing for gauge theories in the continuum limit, since configurations become locally smooth, and thus the interpolation of coarse gauge fields performed prior to the rethermalization step becomes increasingly accurate \cite{Endres:2015yca}.

In this work, we provide numerical evidence that corroborates the heuristic argument that the coupling to the slowest mode decreases with lattice spacing for the case of pure $SU(3)$ gauge theory.
However, we find that the coupling does not decrease at the exponential rate needed to realize an improved rethermalization exponent for lattice spacings in the regime $a\in [0.02,0.06]$ fm that we study.
Rather, the coupling diminishes approximately quadratically with lattice spacing.
Despite this finding, the numerical results suggest that there exists a lattice spacing beyond which the unthermalized bias associated with excited modes of the Markov process becomes negligible for a given desired level of statistics.

\section{Methods and Results}

We study the scaling behavior of the rethermalization timescale, as probed by the average Yang-Mills action density $E(t)$ evaluated at a Wilson flow time $t=w_{0.4}^2/4$, for four prolongated ultrafine ensembles; following Ref. \cite{Borsanyi:2012zs}, the scale $w_{0.4}$ is defined by
\begin{eqnarray}
\left. t \frac{\textrm{d}}{\textrm{d}t} t^2 E(t)  \right|_{t=w_{0.4}^2} = 0.4\ .
\end{eqnarray}
The targeted fine ensembles correspond to a single fixed physical volume of $1.92$ fm, and the lattice spacings $0.06$, $0.04$, $0.03$, and $0.02$ fm.
These ensembles are initially prepared by interpolating fully decorrelated coarse ensembles that have been generated using nonperturbatively matched coarse actions.
All coarse and fine ensembles were generated using the Wilson gauge action; ensemble parameters used for this study are summarized in \Tab{ensemble_param}.

The ensembles were generated using a combination of MCMC algorithms as follows.
$16^3$, $24^3$ and $32^3$ ensembles were initially generated using the Cabibbo-Marinari HB algorithm \cite{CABIBBO1982387} with ten over-relaxation sweeps \cite{PhysRevLett.58.2394} per HB update (one update attempt per link per sweep).
These configurations were subsequently prolongated and rethermalized to produce decorrelated fine $32^2$, $48^3$, and $64^3$ ensembles.
The $48^3$ ensemble was once again prolongated and rethermalized to produced a decorrelated fine $96^3$ ensemble.
In all cases, prolongation of the coarse configurations was performed by gauge field interpolation \cite{Luscher:1981zq,Phillips:1986qd,'tHooft1995491}, following the staged approach described in Ref. \cite{Endres:2015yca}.
Rethermalization of all prolongated ensembles was performed using the HB algorithm (ten update attempts per link per sweep) without over-relaxation.

\begin{table}
\caption{%
\label{tab:ensemble_param}%
Ensemble parameters considered in this work.
Note that the lattice spacings provided below are approximate, and based on numerical estimates for $w_{0.4}/a$, the physical value of the Sommer scale (taken to be $r_0 = 0.5$ fm), and the continuum conversion factor between $r_0$ and reference scale $w_{0.4}$ determined in Ref.~\cite{Asakawa:2015vta}.
}
\begin{ruledtabular}
\begin{tabular}{ccc}
$(L/a)^3\times (T/a)$& $\beta$ & $a$ [fm] \\
\hline
$16^3\times 32$    & 5.87793 & 0.120 \\
$24^3\times 48$    & 6.10050 & 0.081 \\
$32^3\times 64$    & 6.30168 & 0.060 \\
$48^3\times 96$    & 6.59773 & 0.041 \\
$64^3\times 128$   & 6.81596 & 0.030 \\
$96^3\times 192$   & 7.13388 & 0.020 \\
\end{tabular}
\end{ruledtabular}
\end{table}

The coarse and fine actions were consistently matched using the scale parameter $w_{0.4}$, determined in part by using the parametrization with respect to the gauge coupling $\beta$ provided in Ref.~\cite{Asakawa:2015vta}.
Note that that study quotes $0.5\%$ uncertainties in the parametrization over the interval $\beta\in [6.3,7.5]$, and the coarsest two ensembles lie outside this window (for the coarsest ensemble, we performed the nonperturbative tuning independently).
We have validated the scale settings for all but the finest ensemble; results are provided in \Tab{ensemble_matching} and compared with results provided in Ref.~\cite{Asakawa:2015vta}.
For the  ensembles with $L/a=32,48,64$, we find agreement within approximately 2\% of the values predicted by the parametrization of Ref.~\cite{Asakawa:2015vta}, and therefore assume comparable uncertainties in the scale for our finest ensemble, corresponding to $L/a=96$.
In order to make fair comparisons of the rethermalization behavior at different lattice spacings, and ultimately extract reliable scaling properties, it is important to maintain accurate matching between coarse and fine lattices (i.e., between $L/a=16\sim 32$, $L/a=24\sim48$, $L/a=32\sim64$ and $L/a=48\sim96$); here we find agreement in the scale setting parameter to within the quoted statistical uncertainties of \Tab{ensemble_matching}.

\begin{table}
\caption{%
\label{tab:ensemble_matching}%
Scale setting results.
Nominal values for the reference scale determined by Ref.~\cite{Asakawa:2015vta} are indicated by an asterisk.
The configurations used to estimate the scale represent a subset of the total number of configurations generated.
}
\begin{ruledtabular}
\begin{tabular}{ccccc}
$(L/a)^3\times (T/a)$ & no. cfgs. & $w_{0.4}/a$ & $w^*_{0.4}/a$  \\
\hline
$16^3\times 32$    & 700 & 1.605(2)  & 1.733(9)  \\
$24^3\times 48$    & 600 & 2.375(2)  & 2.416(12) \\
$32^3\times 64$    & 400 & 3.218(4)  & 3.221(16) \\
$48^3\times 96$    & 100 & 4.739(11) & 4.832(24) \\
$64^3\times 128$   & 10  & 6.48(5)   & 6.442(32) \\
$96^3\times 192$   & -   & -         & 9.664(48) \\
\end{tabular}
\end{ruledtabular}
\end{table}

The prolongated configurations were evolved toward equilibrium using a total of $\tau_\textrm{max}$ MCMC updates (we validate the choice of $\tau_\textrm{max}$ in a subsequent analysis by demanding that $\tau_\textrm{max}$ be greater than the observed rethermalization time by a factor of approximately 5-10), with intermediate measurements of the Wilson flowed plaquette action density.
Uncorrelated measurements were performed at rethermalization times $\lfloor \hat \tau_n +1/2 \rfloor$, where
\begin{eqnarray}
\hat \tau_n =  - \frac{1}{\hat m} \log\left( 1-\frac{n}{N_\textrm{meas}+1} \right)\ ,
\end{eqnarray}
\begin{eqnarray}
\hat m = -\frac{1}{\tau_\textrm{max}} \log\left( \frac{1}{N_\textrm{meas}+1} \right)\ ,
\end{eqnarray}
and $n=1,\ldots,N_\textrm{meas}$.
Each measurement was performed using disjoint subsets of configurations, each of size $M_\textrm{cfg}$ drawn from the given ensemble.
The values of $\tau_\textrm{max}$, $N_\textrm{meas}$ and $M_\textrm{cfg}$ are provided in \Tab{retherm_param} for each ensemble considered; note that in each case the total number of decorrelated configurations generated by the end of the rethermalization is $N_\textrm{cfg} = N_\textrm{meas} M_\textrm{cfg}$.
Between $N_\textrm{meas}=20-70$ uncorrelated measurements were made in total, with each measurement performed on ensembles of size $M_\textrm{cfg} = 10$ (9 for the $L/a=96$ ensemble).

\begin{table}
\caption{%
\label{tab:retherm_param}%
Rethermalization measurement parameters.
}
\begin{ruledtabular}
\begin{tabular}{cccc}
$(L/a)^3\times (T/a)$& $\tau_\textrm{max}$ & $N_\textrm{meas}$ & $M_\textrm{cfg}$ \\
\hline
$32^3\times 64$    & 275  & 70  & 10\\
$48^3\times 96$    & 500  & 40  & 10\\
$64^3\times 128$   & 750  & 20  & 10\\
$96^3\times 192$   & 1000 & 20  & 9 \\
\end{tabular}
\end{ruledtabular}
\end{table}

\begin{table}
\caption{%
\label{tab:retherm_fit_param}%
Rethermalization fit results.
}
\begin{ruledtabular}
\begin{tabular}{cccccc}
$(L/a)^3\times (T/a)$ & $c_0$ & $\log_{10} c_R$ & $\log_{10}\tau_R$ & $\chi^2/\textrm{d.o.f.}$ \\
\hline
$32^3\times 64$   & 0.1259(1)  & -1.71(1) & 1.21(2)  & 2.1 \\
$48^3\times 96$   & 0.1284(2)  & -2.08(2) & 1.58(5)  & 1.6 \\
$64^3\times 128$  & 0.1275(2)  & -2.18(5) & 1.72(11) & 1.1 \\
$96^3\times 192$  & 0.1289(6)  & -2.61(9) & 2.23(29) & 0.6 \\
\end{tabular}
\end{ruledtabular}
\end{table}

\begin{figure}
\includegraphics[width=0.47\textwidth]{\figdir/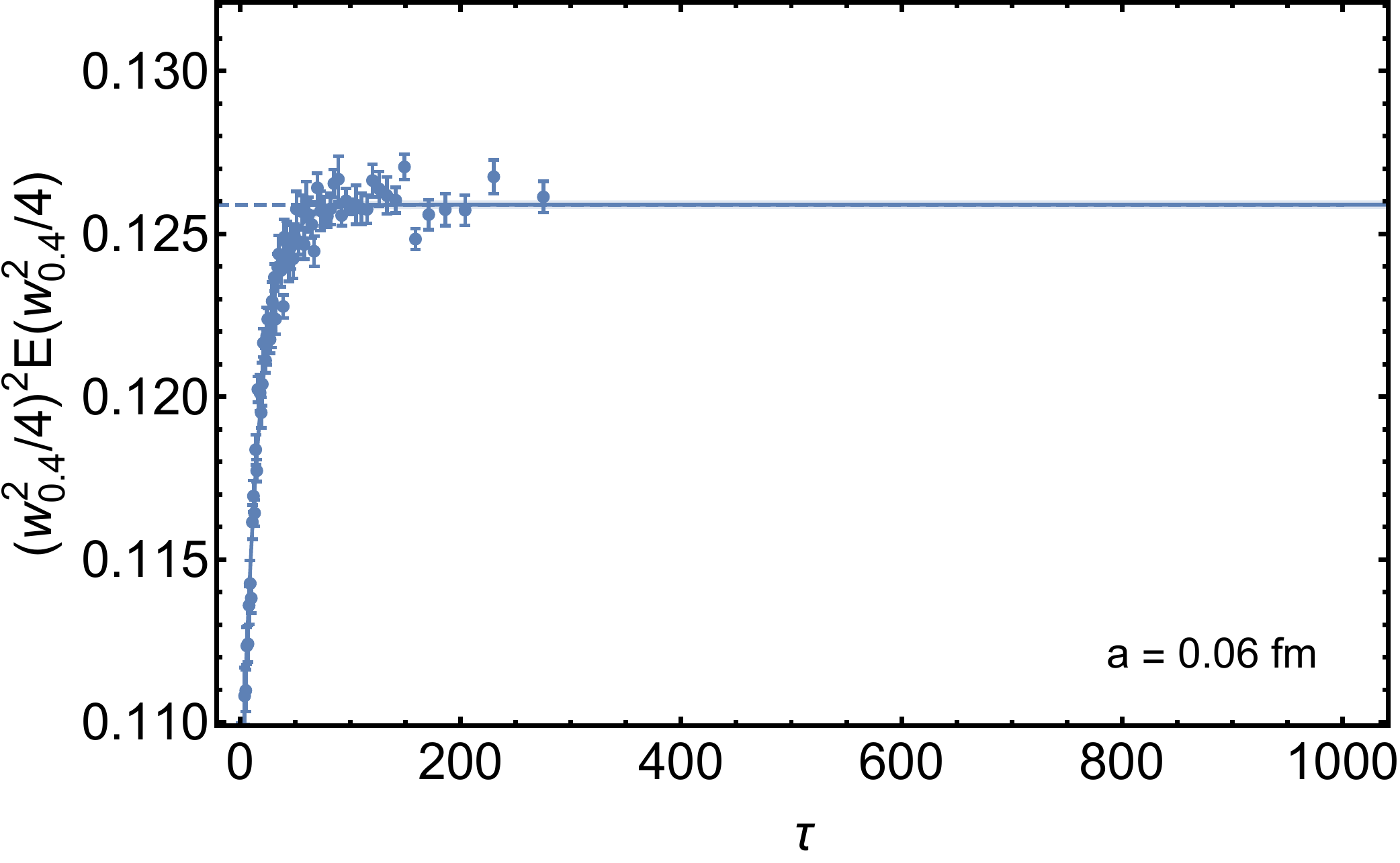}
\includegraphics[width=0.47\textwidth]{\figdir/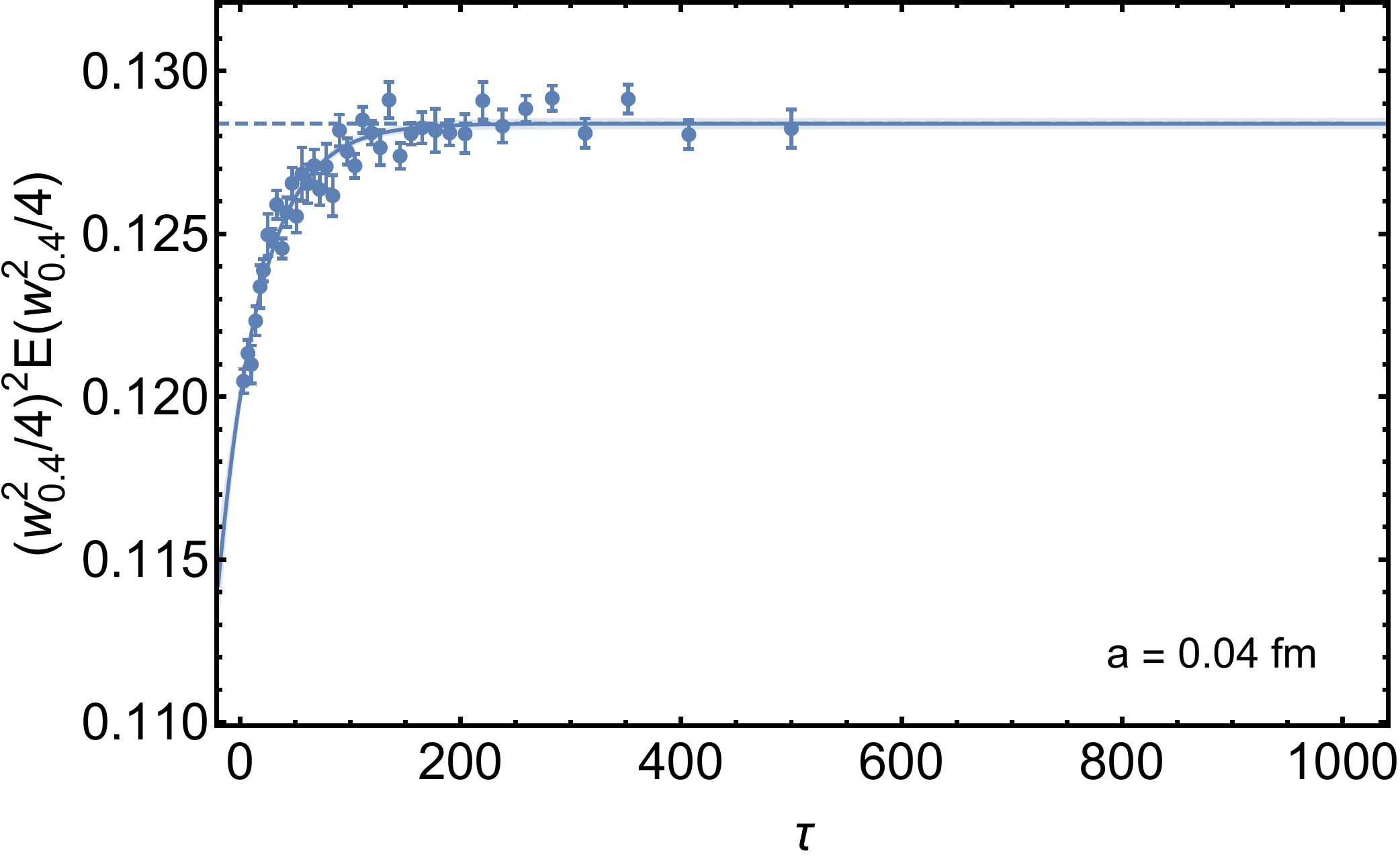}
\includegraphics[width=0.47\textwidth]{\figdir/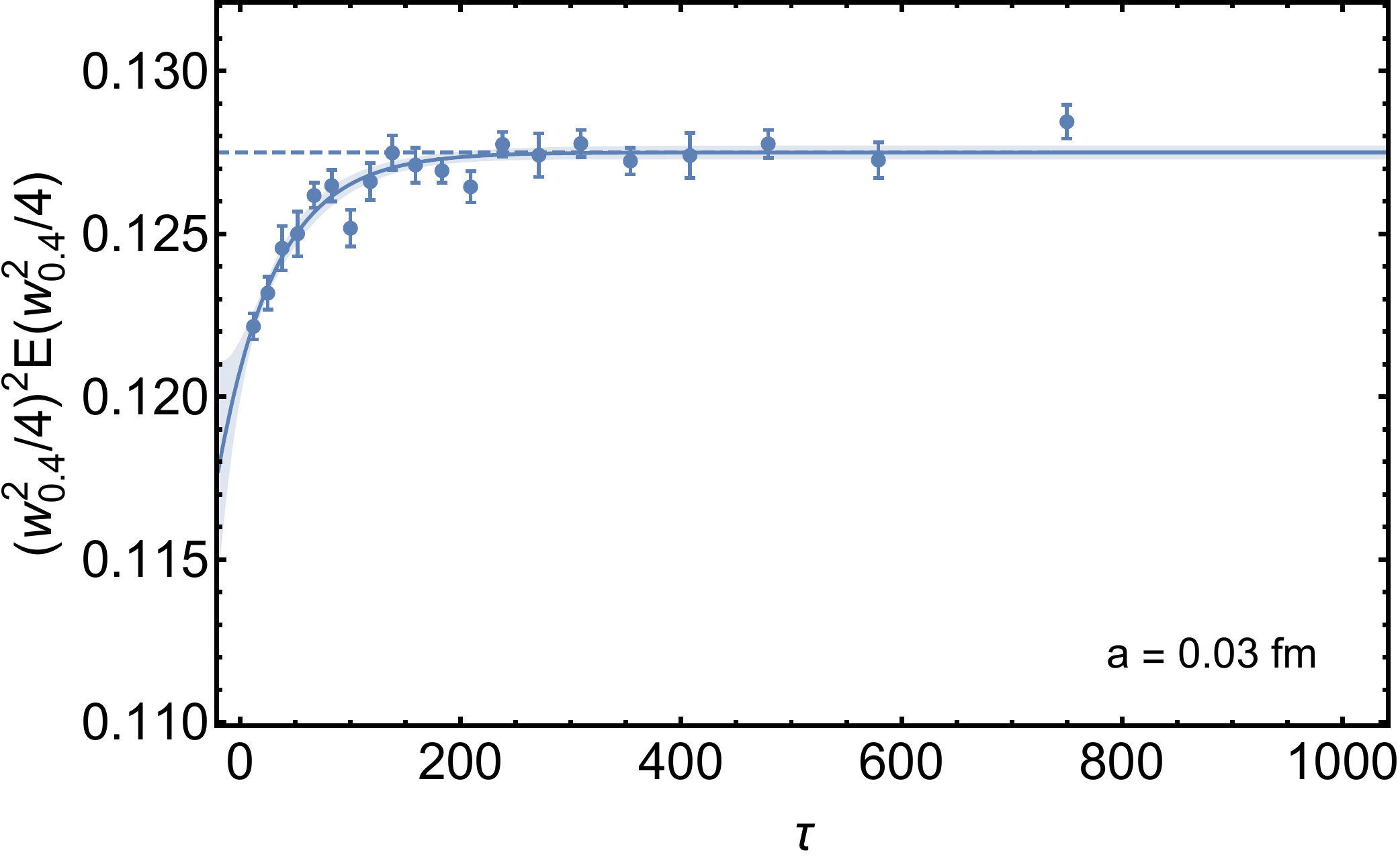}
\includegraphics[width=0.47\textwidth]{\figdir/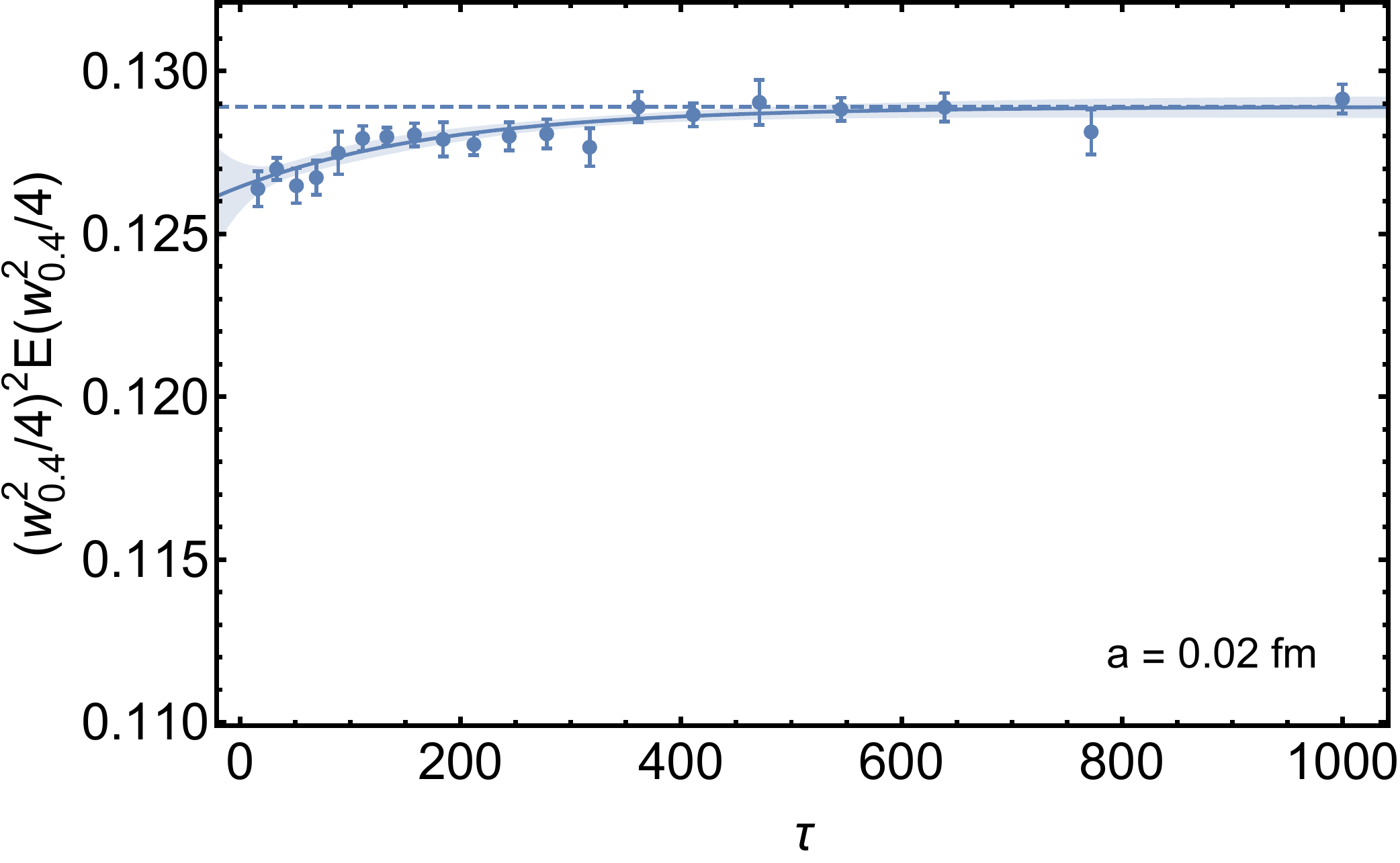}
\caption{\label{fig:fig1}%
Rethermalization time dependence of the Yang-Mills action density at flow time $w_{0.4}^2/4$ for $a=0.06$ fm, $a=0.04$ fm, $a=0.03$ fm, and $a=0.02$ fm.
Note that each data point is statistically independent and the error bands are for uncorrelated exponential fits to the data ($\chi^2/\textrm{d.o.f.}$ range: $0.6-2.1$).
}
\end{figure}

In order to understand the rethermalization time dependence of our prolongated ensembles, we modeled the long-distance observable $\langle {\cal O} \rangle \equiv (w_{0.4}^2/4)^2 E(w^2_{0.4}/4)$ by a single-exponential fit function of the form
\begin{eqnarray}
\langle {\cal O}\rangle = c_0 - c_R e^{-\tau/\tau_R}\ ,
\end{eqnarray}
where $c_0$, $c_R$, and $\tau_R$ are fit parameters and $\tau$ represents the number of MCMC updates.
Least-squares fits were performed using ensemble estimates of $\langle {\cal O}\rangle$ taken over the entire MCMC time range (i.e., $\lfloor \hat \tau_n +1/2 \rfloor$ for $n=1,\ldots,N_\textrm{meas}$); fit results are provided in \Tab{retherm_fit_param} and shown in \Fig{fig1}, along with estimates of the action density measured at each rethermalization time.
Note that the fit values obtained for $c_0$ correspond to a $\tau\to\infty$ extrapolation of the estimator for ${\cal O}$, and are expected to exhibit small variations due to lattice spacing and Wilson flow step size artifacts.
The goodness of these fits, as characterized by the chi-squared per degree of freedom ($\chi^2/\textrm{d.o.f.}$), range from 0.6 to 2.1 and indicate that the model provides an acceptable description of the data.
In particular, except for perhaps the coarsest ensemble, there is little statistically meaningful evidence for exponential contamination beyond the leading ``excited state.'' 

\begin{figure}
\includegraphics[width=0.47\textwidth]{\figdir/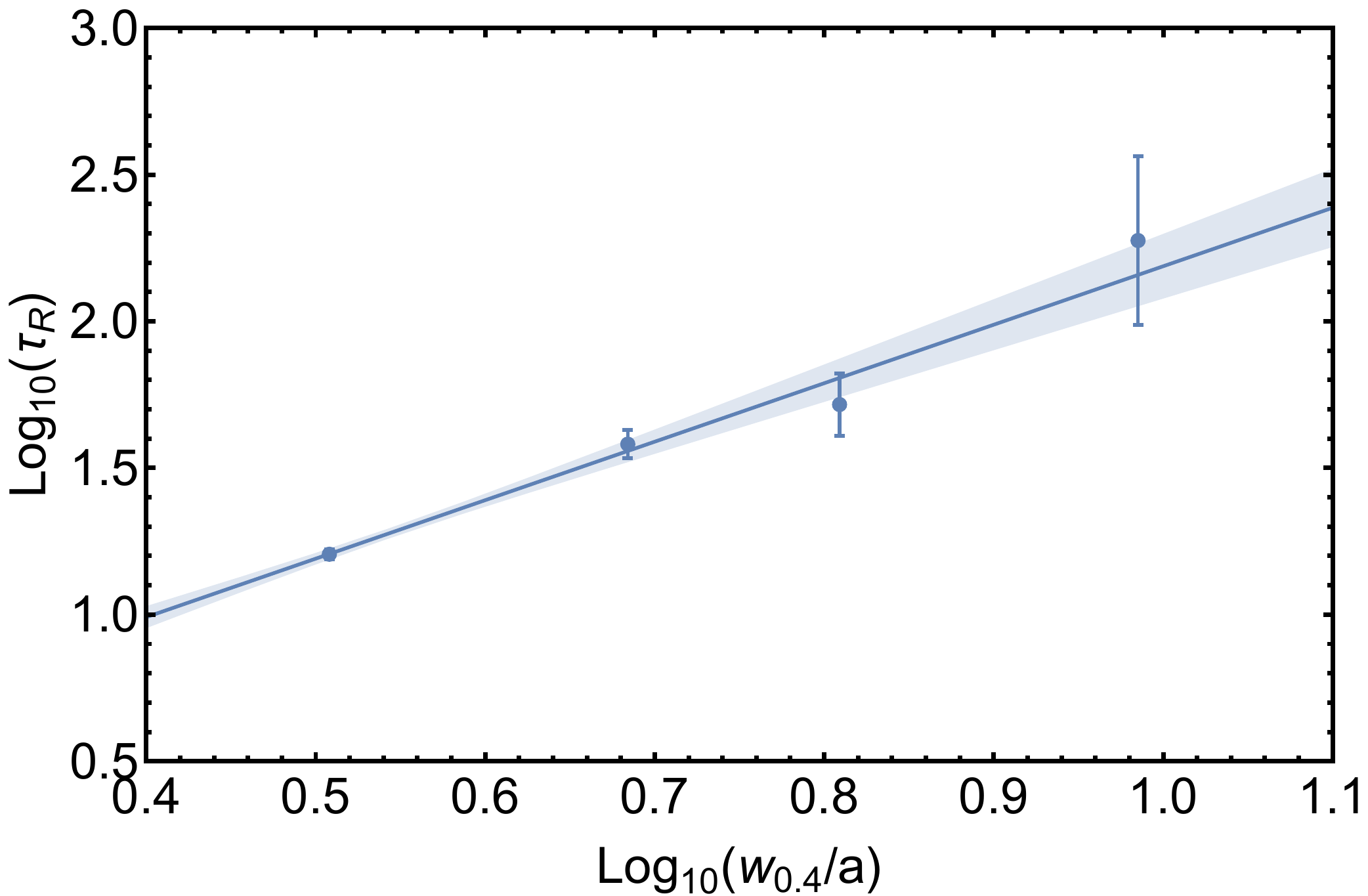}
\includegraphics[width=0.47\textwidth]{\figdir/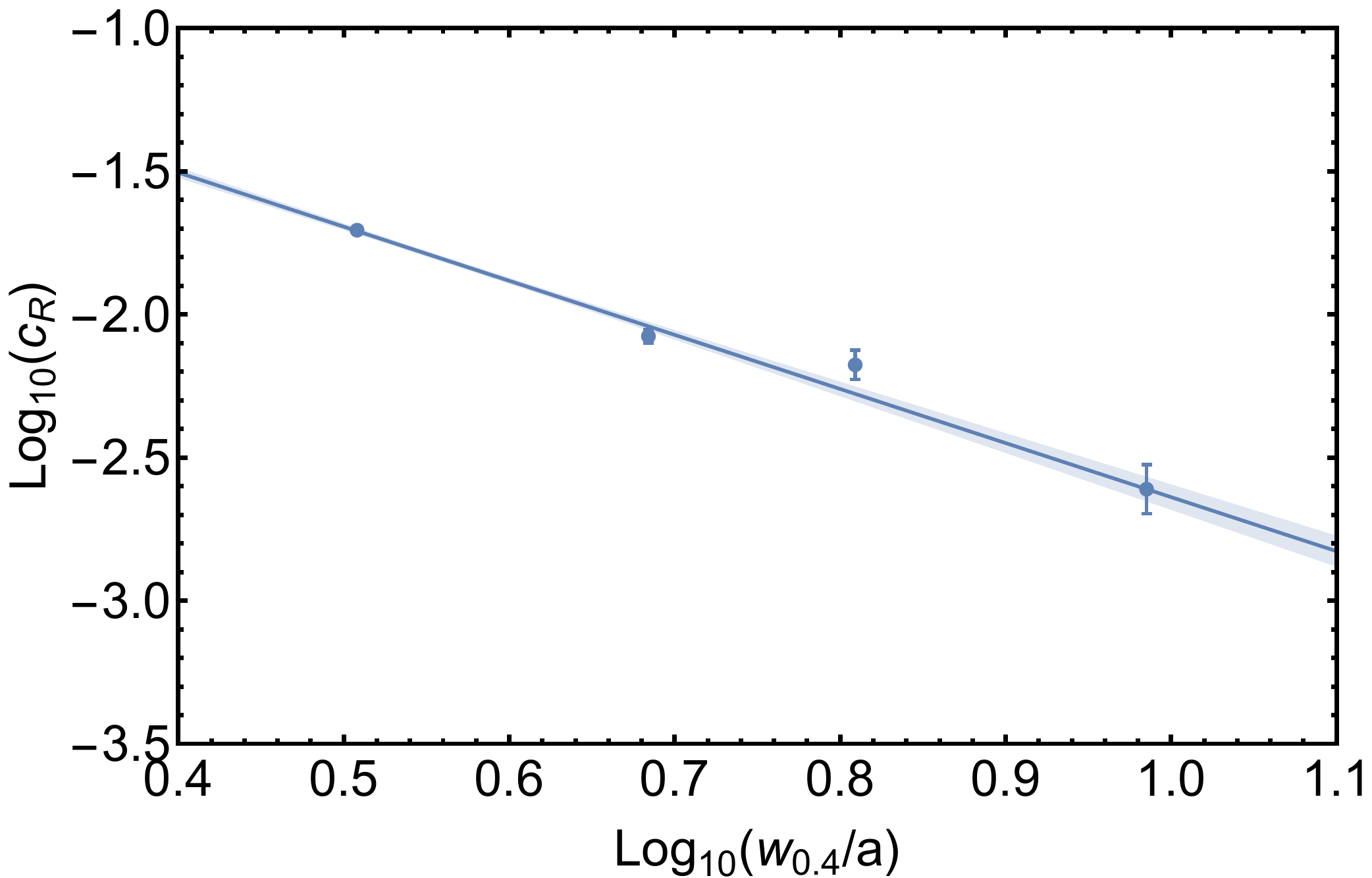}
\caption{\label{fig:fig2}%
Lattice spacing dependence of the rethermalization timescale ($\tau_R$) and coupling ($c_R$), determined from single-exponential fits to the rethermalization time dependence of the action density at flow time $w_{0.4}^2/4$.
The solid line and error band indicate a linear fit to all available data points.
}
\end{figure}

Given the numerical validity of the single-exponential fits, the fit coefficients may be identified with the timescale $\tau_R = \tau_n$ and product of overlap factors
\begin{eqnarray}
c_R = c_n \equiv \langle {\cal O} | \chi_n \rangle \langle \tilde\chi_n | {\cal P}_\textrm{prolongated} \rangle \ ,
\label{eq:overlap}
\end{eqnarray}
associated with the Markov process for {\it some} mode $n>0$ (or some combination of modes with timescales too close to resolve), where the latter provides indirect insight into the strength of the coupling between the prolongated configuration distribution and excited modes of evolution.
\Fig{fig2} shows a plot of the lattice spacing dependence of the extracted rethermalization times, $\tau_R$ and overlap factors, $c_R$.
A fit to these data yields
\begin{eqnarray}
\log_{10} \tau_R = 0.19(11) + 1.99(19) \log_{10}\frac{w_{0.4}}{a}\ ,
\end{eqnarray}
and
\begin{eqnarray}
\log_{10} c_R = -0.749(53) - 1.89(9) \log_{10}\frac{w_{0.4}}{a}\ ,
\label{eq:c1fit}
\end{eqnarray}
respectively.
From the observed lattice spacing dependence of $\tau_R$, the rethermalization scaling exponent appears consistent with $z_\textrm{R}=2$, which is typical for diffusive processes.
Although the coupling strength, $c_R$, appears to diminish quadratically with the lattice spacing, disambiguating the lattice spacing dependence of the individual overlap factors appearing in \Eq{overlap} is not possible given the present data.
However, under the reasonable assumption that the overlap factor $\langle {\cal O} | \chi_n \rangle$ has a nonzero continuum limit\footnote{If for $n>0$ one assumes $\langle {\cal O} | \chi_n \rangle$ diminishes with lattice spacing for suitably finite renormalized operators, ${\cal O}$, then in the continuum limit, reliable estimates of ${\cal O}$ would be possible for an ensemble drawn from an {\it arbitrary} distribution, since under this assumption $c_n \to 0$ independently of the initial probability distribution.
This cannot be the case for, for example, an initial free-field probability distribution, and therefore one arrives at a contradiction.} and the same mode is dominant at each lattice spacing, we find clear evidence of decoupling of the prolongated distribution from the dominant (nontopological) mode in the continuum limit.

\section{Discussion}

In previous studies of both short- and long-distance observables in pure $SU(3)$ gauge theory \cite{Endres:2015yca} and $SU(2)$ gauge theory with heavy dynamical fermions \cite{Detmold:2016rnh}, it was established that a properly matched and prolongated coarse ensemble can yield vanishing overlap onto the slowest nontopological modes of the Markov process for both HB and HMC algorithms.
In this study, based on the quality of our single-exponential fits to the long-distance observable $\langle \calO \rangle$ with respect to the rethermalization time, we find little evidence for {\it any} but a single, approximately diffusive, mode of evolution (hereafter referred to as a ``rethermalization mode'') contributing to the rethermalization of prolongated ensembles.
The degrees of freedom of the prolongated configurations that incorrectly describe the target fine probability distribution dominantly couple to the rethermalization mode, but with a strength that diminishes quadratically with the lattice spacing.
For a given level of statistical precision, the findings suggest that at a sufficiently fine lattice spacing rethermalization may become unnecessary.

Specifically, even in the regime $\tau\ll\tau_R$, and for a given fixed level of statistical precision, the bias associated with the rethermalization mode contamination is negligible provided $c_R \alt \sigma/\sqrt{N_\textrm{conf}}$, where $\sigma$ is the standard deviation of the measured observable distribution, and $N_{\rm conf}$ is the number of statistically independent configurations used in the measurement.
In the case of the action density at flow time $w_{0.4}^2/4$, the fitted form in Eq.~(\ref{eq:c1fit}) predicts this to occur at a lattice spacing $a\sim 0.01$ fm (20 GeV) at the present level of statistics.
Since $c_R \sim a^2$, maintaining this condition at a higher level of statistics demands that the lattice spacing be reduced, scaling like $a\sim {N_\textrm{conf}^{-1/4}}$ for a fixed physical volume.
This observation and the corresponding scale at which rethermalization becomes unnecessary may be specific to the operator being studied.
However, our previous study \cite{Endres:2015yca} has shown rethermalization times are relatively insensitive over a large class of long-distance operators.

Further suppression of the bias associated with rethermalization mode contamination may be possible by improving the scaling properties of rethermalization mode coupling to the prolongated distribution.
It is not clear, however, whether the quadratic dependence of $c_R$ observed in \Eq{c1fit} arises out of geometrical considerations (noting that the current interpolation scheme involves mapping coarse configurations to a fine grid and then locally minimizing the Wilson action with respect to the remaining gauge links, the interpolation from coarse to fine is expected to only be valid up to quadratic corrections in the lattice spacing) or some other source of lattice spacing dependence (e.g., the quality and consistency of the matching at different scales).
If the origin for this scaling is indeed geometrical, then a higher order prolongation scheme could yield a higher order dependence of $c_R$ on the lattice spacing.

Focusing on interpolation-based prolongation operations, and assuming the renormalization-group matching produces coarse links that are properly sampled from the standpoint of the fine action, then the most probable values for the remaining fine links are the ones that minimized the classical continuum equations of motion--up to a statistical variance that must diminish with lattice spacing.
This suggests that a higher order interpolation scheme might involve minimizing a classically improved lattice gauge action, subject to the constraint that coarse links be held fixed.
A practical realization for such a scheme might involve two passes: in the first pass, an interpolation is performed locally using the lower order scheme as in this study, and in the subsequent pass, a global minimization of the improved gauge action is performed subject to the coarse link constraints.
Use of the lower order interpolated gauge links as a starting point (e.g., in an iterative optimization scheme) would help to ensure that the desired global minimum is reached.

Whether or not such an improved interpolation scheme yields an improved prolongation operation, and whether or not an improved prolongation operation yields improved scaling of the overlap of the resulting configuration distribution onto the observed rethermalization mode remain interesting and open questions.
Regardless of the outcome, and given the observed decrease in contamination from unwanted rethermalization modes, use of prolongated coarse, matched ensembles to rapidly thermalize fine ensemble streams with well-sampled topology appears to be increasingly feasible as the continuum limit is approached.

\begin{acknowledgments}
We would like to thank Richard Brower, Zohreh Davoudi, Kostas Orginos, Andrew Pochinsky, and Phiala Shanahan for informative discussions.
All simulations were performed using a modified version of the Chroma Software System for lattice QCD~\cite{Edwards:2004sx}.
Computations for this study were carried out on facilities of the USQCD Collaboration, which are funded by the Office of Science of the U.~S.~Department of Energy. 
This work was partially supported by the U.~S.~Department of Energy through Early Career Research Award No. DE-SC0010495 and Grant No. DE-SC0011090 and by the SciDAC4 Grant No. DE-SC0018121.
\end{acknowledgments}

\bibliography{mstherm_scaling}

\begin{thebibliography}{33}%
\makeatletter
\providecommand \@ifxundefined [1]{%
 \@ifx{#1\undefined}
}%
\providecommand \@ifnum [1]{%
 \ifnum #1\expandafter \@firstoftwo
 \else \expandafter \@secondoftwo
 \fi
}%
\providecommand \@ifx [1]{%
 \ifx #1\expandafter \@firstoftwo
 \else \expandafter \@secondoftwo
 \fi
}%
\providecommand \natexlab [1]{#1}%
\providecommand \enquote  [1]{``#1''}%
\providecommand \bibnamefont  [1]{#1}%
\providecommand \bibfnamefont [1]{#1}%
\providecommand \citenamefont [1]{#1}%
\providecommand \href@noop [0]{\@secondoftwo}%
\providecommand \href [0]{\begingroup \@sanitize@url \@href}%
\providecommand \@href[1]{\@@startlink{#1}\@@href}%
\providecommand \@@href[1]{\endgroup#1\@@endlink}%
\providecommand \@sanitize@url [0]{\catcode `\\12\catcode `\$12\catcode
  `\&12\catcode `\#12\catcode `\^12\catcode `\_12\catcode `\%12\relax}%
\providecommand \@@startlink[1]{}%
\providecommand \@@endlink[0]{}%
\providecommand \url  [0]{\begingroup\@sanitize@url \@url }%
\providecommand \@url [1]{\endgroup\@href {#1}{\urlprefix }}%
\providecommand \urlprefix  [0]{URL }%
\providecommand \Eprint [0]{\href }%
\providecommand \doibase [0]{http://dx.doi.org/}%
\providecommand \selectlanguage [0]{\@gobble}%
\providecommand \bibinfo  [0]{\@secondoftwo}%
\providecommand \bibfield  [0]{\@secondoftwo}%
\providecommand \translation [1]{[#1]}%
\providecommand \BibitemOpen [0]{}%
\providecommand \bibitemStop [0]{}%
\providecommand \bibitemNoStop [0]{.\EOS\space}%
\providecommand \EOS [0]{\spacefactor3000\relax}%
\providecommand \BibitemShut  [1]{\csname bibitem#1\endcsname}%
\let\auto@bib@innerbib\@empty
\bibitem [{\citenamefont {Babich}\ \emph {et~al.}(2010)\citenamefont {Babich},
  \citenamefont {Brannick}, \citenamefont {Brower}, \citenamefont {Clark},
  \citenamefont {Manteuffel}, \citenamefont {McCormick}, \citenamefont
  {Osborn},\ and\ \citenamefont {Rebbi}}]{Babich:2010qb}%
  \BibitemOpen
  \bibfield  {author} {\bibinfo {author} {\bibfnamefont {R.}~\bibnamefont
  {Babich}}, \bibinfo {author} {\bibfnamefont {J.}~\bibnamefont {Brannick}},
  \bibinfo {author} {\bibfnamefont {R.~C.}\ \bibnamefont {Brower}}, \bibinfo
  {author} {\bibfnamefont {M.~A.}\ \bibnamefont {Clark}}, \bibinfo {author}
  {\bibfnamefont {T.~A.}\ \bibnamefont {Manteuffel}}, \bibinfo {author}
  {\bibfnamefont {S.~F.}\ \bibnamefont {McCormick}}, \bibinfo {author}
  {\bibfnamefont {J.~C.}\ \bibnamefont {Osborn}}, \ and\ \bibinfo {author}
  {\bibfnamefont {C.}~\bibnamefont {Rebbi}},\ }\href {\doibase
  10.1103/PhysRevLett.105.201602} {\bibfield  {journal} {\bibinfo  {journal}
  {Phys. Rev. Lett.}\ }\textbf {\bibinfo {volume} {105}},\ \bibinfo {pages}
  {201602} (\bibinfo {year} {2010})},\ \Eprint {http://arxiv.org/abs/1005.3043}
  {arXiv:1005.3043 [hep-lat]} \BibitemShut {NoStop}%
\bibitem [{\citenamefont {Babich}\ \emph {et~al.}(2009)\citenamefont {Babich},
  \citenamefont {Brannick}, \citenamefont {Brower}, \citenamefont {Clark},
  \citenamefont {Cohen}, \citenamefont {Osborn},\ and\ \citenamefont
  {Rebbi}}]{Babich:2009pc}%
  \BibitemOpen
  \bibfield  {author} {\bibinfo {author} {\bibfnamefont {R.}~\bibnamefont
  {Babich}}, \bibinfo {author} {\bibfnamefont {J.}~\bibnamefont {Brannick}},
  \bibinfo {author} {\bibfnamefont {R.~C.}\ \bibnamefont {Brower}}, \bibinfo
  {author} {\bibfnamefont {M.~A.}\ \bibnamefont {Clark}}, \bibinfo {author}
  {\bibfnamefont {S.~D.}\ \bibnamefont {Cohen}}, \bibinfo {author}
  {\bibfnamefont {J.~C.}\ \bibnamefont {Osborn}}, \ and\ \bibinfo {author}
  {\bibfnamefont {C.}~\bibnamefont {Rebbi}},\ }\bibfield  {booktitle} {\emph
  {\bibinfo {booktitle} {{Proceedings, 27th International Symposium on Lattice
  field theory (Lattice 2009)}}},\ }\href@noop {} {\bibfield  {journal}
  {\bibinfo  {journal} {PoS}\ }\textbf {\bibinfo {volume} {LAT2009}},\ \bibinfo
  {pages} {031} (\bibinfo {year} {2009})},\ \Eprint
  {http://arxiv.org/abs/0912.2186} {arXiv:0912.2186 [hep-lat]} \BibitemShut
  {NoStop}%
\bibitem [{\citenamefont {Frommer}\ \emph {et~al.}(2014)\citenamefont
  {Frommer}, \citenamefont {Kahl}, \citenamefont {Krieg}, \citenamefont
  {Leder},\ and\ \citenamefont {Rottmann}}]{Frommer:2013fsa}%
  \BibitemOpen
  \bibfield  {author} {\bibinfo {author} {\bibfnamefont {A.}~\bibnamefont
  {Frommer}}, \bibinfo {author} {\bibfnamefont {K.}~\bibnamefont {Kahl}},
  \bibinfo {author} {\bibfnamefont {S.}~\bibnamefont {Krieg}}, \bibinfo
  {author} {\bibfnamefont {B.}~\bibnamefont {Leder}}, \ and\ \bibinfo {author}
  {\bibfnamefont {M.}~\bibnamefont {Rottmann}},\ }\href {\doibase
  10.1137/130919507} {\bibfield  {journal} {\bibinfo  {journal} {SIAM J. Sci.
  Comput.}\ }\textbf {\bibinfo {volume} {36}},\ \bibinfo {pages} {A1581}
  (\bibinfo {year} {2014})},\ \Eprint {http://arxiv.org/abs/1303.1377}
  {arXiv:1303.1377 [hep-lat]} \BibitemShut {NoStop}%
\bibitem [{\citenamefont {Brannick}\ \emph {et~al.}(2016)\citenamefont
  {Brannick}, \citenamefont {Frommer}, \citenamefont {Kahl}, \citenamefont
  {Leder}, \citenamefont {Rottmann},\ and\ \citenamefont
  {Strebel}}]{Brannick:2014vda}%
  \BibitemOpen
  \bibfield  {author} {\bibinfo {author} {\bibfnamefont {J.}~\bibnamefont
  {Brannick}}, \bibinfo {author} {\bibfnamefont {A.}~\bibnamefont {Frommer}},
  \bibinfo {author} {\bibfnamefont {K.}~\bibnamefont {Kahl}}, \bibinfo {author}
  {\bibfnamefont {B.}~\bibnamefont {Leder}}, \bibinfo {author} {\bibfnamefont
  {M.}~\bibnamefont {Rottmann}}, \ and\ \bibinfo {author} {\bibfnamefont
  {A.}~\bibnamefont {Strebel}},\ }\href {\doibase 10.1007/s00211-015-0725-6}
  {\bibfield  {journal} {\bibinfo  {journal} {Numer. Math.}\ }\textbf {\bibinfo
  {volume} {132}},\ \bibinfo {pages} {463} (\bibinfo {year} {2016})},\ \Eprint
  {http://arxiv.org/abs/1410.7170} {arXiv:1410.7170 [hep-lat]} \BibitemShut
  {NoStop}%
\bibitem [{\citenamefont {L{\"u}scher}\ and\ \citenamefont
  {Weisz}(2001)}]{Luscher:2001up}%
  \BibitemOpen
  \bibfield  {author} {\bibinfo {author} {\bibfnamefont {M.}~\bibnamefont
  {L{\"u}scher}}\ and\ \bibinfo {author} {\bibfnamefont {P.}~\bibnamefont
  {Weisz}},\ }\href {\doibase 10.1088/1126-6708/2001/09/010} {\bibfield
  {journal} {\bibinfo  {journal} {JHEP}\ }\textbf {\bibinfo {volume} {09}},\
  \bibinfo {pages} {010} (\bibinfo {year} {2001})},\ \Eprint
  {http://arxiv.org/abs/hep-lat/0108014} {arXiv:hep-lat/0108014 [hep-lat]}
  \BibitemShut {NoStop}%
\bibitem [{\citenamefont {Ce}\ \emph {et~al.}(2016)\citenamefont {Ce},
  \citenamefont {Giusti},\ and\ \citenamefont {Schaefer}}]{Ce:2016idq}%
  \BibitemOpen
  \bibfield  {author} {\bibinfo {author} {\bibfnamefont {M.}~\bibnamefont
  {Ce}}, \bibinfo {author} {\bibfnamefont {L.}~\bibnamefont {Giusti}}, \ and\
  \bibinfo {author} {\bibfnamefont {S.}~\bibnamefont {Schaefer}},\ }\href
  {\doibase 10.1103/PhysRevD.93.094507} {\bibfield  {journal} {\bibinfo
  {journal} {Phys. Rev.}\ }\textbf {\bibinfo {volume} {D93}},\ \bibinfo {pages}
  {094507} (\bibinfo {year} {2016})},\ \Eprint
  {http://arxiv.org/abs/1601.04587} {arXiv:1601.04587 [hep-lat]} \BibitemShut
  {NoStop}%
\bibitem [{\citenamefont {García~Vera}\ and\ \citenamefont
  {Schaefer}(2016)}]{Vera:2016xpp}%
  \BibitemOpen
  \bibfield  {author} {\bibinfo {author} {\bibfnamefont {M.}~\bibnamefont
  {García~Vera}}\ and\ \bibinfo {author} {\bibfnamefont {S.}~\bibnamefont
  {Schaefer}},\ }\href {\doibase 10.1103/PhysRevD.93.074502} {\bibfield
  {journal} {\bibinfo  {journal} {Phys. Rev.}\ }\textbf {\bibinfo {volume}
  {D93}},\ \bibinfo {pages} {074502} (\bibinfo {year} {2016})},\ \Eprint
  {http://arxiv.org/abs/1601.07155} {arXiv:1601.07155 [hep-lat]} \BibitemShut
  {NoStop}%
\bibitem [{\citenamefont {Ce}\ \emph {et~al.}(2017)\citenamefont {Ce},
  \citenamefont {Giusti},\ and\ \citenamefont {Schaefer}}]{Ce:2016ajy}%
  \BibitemOpen
  \bibfield  {author} {\bibinfo {author} {\bibfnamefont {M.}~\bibnamefont
  {Ce}}, \bibinfo {author} {\bibfnamefont {L.}~\bibnamefont {Giusti}}, \ and\
  \bibinfo {author} {\bibfnamefont {S.}~\bibnamefont {Schaefer}},\ }\href
  {\doibase 10.1103/PhysRevD.95.034503} {\bibfield  {journal} {\bibinfo
  {journal} {Phys. Rev.}\ }\textbf {\bibinfo {volume} {D95}},\ \bibinfo {pages}
  {034503} (\bibinfo {year} {2017})},\ \Eprint
  {http://arxiv.org/abs/1609.02419} {arXiv:1609.02419 [hep-lat]} \BibitemShut
  {NoStop}%
\bibitem [{\citenamefont {Goodman}\ and\ \citenamefont
  {Sokal}(1986)}]{Goodman:1986pv}%
  \BibitemOpen
  \bibfield  {author} {\bibinfo {author} {\bibfnamefont {J.}~\bibnamefont
  {Goodman}}\ and\ \bibinfo {author} {\bibfnamefont {A.~D.}\ \bibnamefont
  {Sokal}},\ }\href {\doibase 10.1103/PhysRevLett.56.1015} {\bibfield
  {journal} {\bibinfo  {journal} {Phys. Rev. Lett.}\ }\textbf {\bibinfo
  {volume} {56}},\ \bibinfo {pages} {1015} (\bibinfo {year}
  {1986})}\BibitemShut {NoStop}%
\bibitem [{\citenamefont {Edwards}\ \emph {et~al.}(1991)\citenamefont
  {Edwards}, \citenamefont {Goodman},\ and\ \citenamefont
  {Sokal}}]{Edwards:1990hu}%
  \BibitemOpen
  \bibfield  {author} {\bibinfo {author} {\bibfnamefont {R.~G.}\ \bibnamefont
  {Edwards}}, \bibinfo {author} {\bibfnamefont {J.}~\bibnamefont {Goodman}}, \
  and\ \bibinfo {author} {\bibfnamefont {A.~D.}\ \bibnamefont {Sokal}},\ }\href
  {\doibase 10.1016/0550-3213(91)90357-4} {\bibfield  {journal} {\bibinfo
  {journal} {Nucl. Phys.}\ }\textbf {\bibinfo {volume} {B354}},\ \bibinfo
  {pages} {289} (\bibinfo {year} {1991})}\BibitemShut {NoStop}%
\bibitem [{\citenamefont {Edwards}\ \emph {et~al.}(1992)\citenamefont
  {Edwards}, \citenamefont {Ferreira}, \citenamefont {Goodman},\ and\
  \citenamefont {Sokal}}]{Edwards:1991eg}%
  \BibitemOpen
  \bibfield  {author} {\bibinfo {author} {\bibfnamefont {R.~G.}\ \bibnamefont
  {Edwards}}, \bibinfo {author} {\bibfnamefont {S.~J.}\ \bibnamefont
  {Ferreira}}, \bibinfo {author} {\bibfnamefont {J.}~\bibnamefont {Goodman}}, \
  and\ \bibinfo {author} {\bibfnamefont {A.~D.}\ \bibnamefont {Sokal}},\ }\href
  {\doibase 10.1016/0550-3213(92)90262-A} {\bibfield  {journal} {\bibinfo
  {journal} {Nucl. Phys.}\ }\textbf {\bibinfo {volume} {B380}},\ \bibinfo
  {pages} {621} (\bibinfo {year} {1992})},\ \Eprint
  {http://arxiv.org/abs/hep-lat/9112002} {arXiv:hep-lat/9112002 [hep-lat]}
  \BibitemShut {NoStop}%
\bibitem [{\citenamefont {Janke}\ and\ \citenamefont
  {Sauer}(1994)}]{Janke:1993et}%
  \BibitemOpen
  \bibfield  {author} {\bibinfo {author} {\bibfnamefont {W.}~\bibnamefont
  {Janke}}\ and\ \bibinfo {author} {\bibfnamefont {T.}~\bibnamefont {Sauer}},\
  }\bibfield  {booktitle} {\emph {\bibinfo {booktitle} {{LATTICE 93: 11th
  International Symposium on Lattice Field Theory Dallas, Texas, October 12-16,
  1993}}},\ }\href {\doibase 10.1016/0920-5632(94)90509-6} {\bibfield
  {journal} {\bibinfo  {journal} {Nucl. Phys. Proc. Suppl.}\ }\textbf {\bibinfo
  {volume} {34}},\ \bibinfo {pages} {771} (\bibinfo {year} {1994})},\ \Eprint
  {http://arxiv.org/abs/hep-lat/9312043} {arXiv:hep-lat/9312043 [hep-lat]}
  \BibitemShut {NoStop}%
\bibitem [{\citenamefont {Grabenstein}\ and\ \citenamefont
  {Mikeska}(1994)}]{Grabenstein:1993nh}%
  \BibitemOpen
  \bibfield  {author} {\bibinfo {author} {\bibfnamefont {M.}~\bibnamefont
  {Grabenstein}}\ and\ \bibinfo {author} {\bibfnamefont {B.}~\bibnamefont
  {Mikeska}},\ }\bibfield  {booktitle} {\emph {\bibinfo {booktitle} {{LATTICE
  93: 11th International Symposium on Lattice Field Theory Dallas, Texas,
  October 12-16, 1993}}},\ }\href {\doibase 10.1016/0920-5632(94)90507-X}
  {\bibfield  {journal} {\bibinfo  {journal} {Nucl. Phys. Proc. Suppl.}\
  }\textbf {\bibinfo {volume} {34}},\ \bibinfo {pages} {765} (\bibinfo {year}
  {1994})},\ \Eprint {http://arxiv.org/abs/hep-lat/9311021}
  {arXiv:hep-lat/9311021 [hep-lat]} \BibitemShut {NoStop}%
\bibitem [{\citenamefont {Grabenstein}\ and\ \citenamefont
  {Pinn}(1994)}]{Grabenstein:1994ze}%
  \BibitemOpen
  \bibfield  {author} {\bibinfo {author} {\bibfnamefont {M.}~\bibnamefont
  {Grabenstein}}\ and\ \bibinfo {author} {\bibfnamefont {K.}~\bibnamefont
  {Pinn}},\ }\href {\doibase 10.1103/PhysRevD.50.6998} {\bibfield  {journal}
  {\bibinfo  {journal} {Phys. Rev.}\ }\textbf {\bibinfo {volume} {D50}},\
  \bibinfo {pages} {6998} (\bibinfo {year} {1994})},\ \Eprint
  {http://arxiv.org/abs/hep-lat/9406013} {arXiv:hep-lat/9406013 [hep-lat]}
  \BibitemShut {NoStop}%
\bibitem [{\citenamefont {Endres}\ \emph {et~al.}(2015)\citenamefont {Endres},
  \citenamefont {Brower}, \citenamefont {Detmold}, \citenamefont {Orginos},\
  and\ \citenamefont {Pochinsky}}]{Endres:2015yca}%
  \BibitemOpen
  \bibfield  {author} {\bibinfo {author} {\bibfnamefont {M.~G.}\ \bibnamefont
  {Endres}}, \bibinfo {author} {\bibfnamefont {R.~C.}\ \bibnamefont {Brower}},
  \bibinfo {author} {\bibfnamefont {W.}~\bibnamefont {Detmold}}, \bibinfo
  {author} {\bibfnamefont {K.}~\bibnamefont {Orginos}}, \ and\ \bibinfo
  {author} {\bibfnamefont {A.~V.}\ \bibnamefont {Pochinsky}},\ }\href {\doibase
  10.1103/PhysRevD.92.114516} {\bibfield  {journal} {\bibinfo  {journal} {Phys.
  Rev.}\ }\textbf {\bibinfo {volume} {D92}},\ \bibinfo {pages} {114516}
  (\bibinfo {year} {2015})},\ \Eprint {http://arxiv.org/abs/1510.04675}
  {arXiv:1510.04675 [hep-lat]} \BibitemShut {NoStop}%
\bibitem [{\citenamefont {Detmold}\ and\ \citenamefont
  {Endres}(2016)}]{Detmold:2016rnh}%
  \BibitemOpen
  \bibfield  {author} {\bibinfo {author} {\bibfnamefont {W.}~\bibnamefont
  {Detmold}}\ and\ \bibinfo {author} {\bibfnamefont {M.~G.}\ \bibnamefont
  {Endres}},\ }\href@noop {} {\bibfield  {journal} {\bibinfo  {journal} {Phys.
  Rev.}\ }\textbf {\bibinfo {volume} {D94}},\ \bibinfo {pages} {114502}
  (\bibinfo {year} {2016})},\ \Eprint {http://arxiv.org/abs/1605.09650}
  {arXiv:1605.09650 [hep-lat]} \BibitemShut {NoStop}%
\bibitem [{\citenamefont {Schaefer}\ \emph {et~al.}(2011)\citenamefont
  {Schaefer}, \citenamefont {Sommer},\ and\ \citenamefont
  {Virotta}}]{Schaefer:2010hu}%
  \BibitemOpen
  \bibfield  {author} {\bibinfo {author} {\bibfnamefont {S.}~\bibnamefont
  {Schaefer}}, \bibinfo {author} {\bibfnamefont {R.}~\bibnamefont {Sommer}}, \
  and\ \bibinfo {author} {\bibfnamefont {F.}~\bibnamefont {Virotta}} (\bibinfo
  {collaboration} {ALPHA}),\ }\href {\doibase 10.1016/j.nuclphysb.2010.11.020}
  {\bibfield  {journal} {\bibinfo  {journal} {Nucl. Phys.}\ }\textbf {\bibinfo
  {volume} {B845}},\ \bibinfo {pages} {93} (\bibinfo {year} {2011})},\ \Eprint
  {http://arxiv.org/abs/1009.5228} {arXiv:1009.5228 [hep-lat]} \BibitemShut
  {NoStop}%
\bibitem [{\citenamefont {L{\"u}scher}\ and\ \citenamefont
  {Schaefer}(2011)}]{Luscher:2011kk}%
  \BibitemOpen
  \bibfield  {author} {\bibinfo {author} {\bibfnamefont {M.}~\bibnamefont
  {L{\"u}scher}}\ and\ \bibinfo {author} {\bibfnamefont {S.}~\bibnamefont
  {Schaefer}},\ }\href {\doibase 10.1007/JHEP07(2011)036} {\bibfield  {journal}
  {\bibinfo  {journal} {JHEP}\ }\textbf {\bibinfo {volume} {07}},\ \bibinfo
  {pages} {036} (\bibinfo {year} {2011})},\ \Eprint
  {http://arxiv.org/abs/1105.4749} {arXiv:1105.4749 [hep-lat]} \BibitemShut
  {NoStop}%
\bibitem [{\citenamefont {Mages}\ \emph {et~al.}(2017)\citenamefont {Mages},
  \citenamefont {Toth}, \citenamefont {Borsanyi}, \citenamefont {Fodor},
  \citenamefont {Katz},\ and\ \citenamefont {Szabo}}]{Mages:2015scv}%
  \BibitemOpen
  \bibfield  {author} {\bibinfo {author} {\bibfnamefont {S.}~\bibnamefont
  {Mages}}, \bibinfo {author} {\bibfnamefont {B.~C.}\ \bibnamefont {Toth}},
  \bibinfo {author} {\bibfnamefont {S.}~\bibnamefont {Borsanyi}}, \bibinfo
  {author} {\bibfnamefont {Z.}~\bibnamefont {Fodor}}, \bibinfo {author}
  {\bibfnamefont {S.~D.}\ \bibnamefont {Katz}}, \ and\ \bibinfo {author}
  {\bibfnamefont {K.~K.}\ \bibnamefont {Szabo}},\ }\href {\doibase
  10.1103/PhysRevD.95.094512} {\bibfield  {journal} {\bibinfo  {journal} {Phys.
  Rev.}\ }\textbf {\bibinfo {volume} {D95}},\ \bibinfo {pages} {094512}
  (\bibinfo {year} {2017})},\ \Eprint {http://arxiv.org/abs/1512.06804}
  {arXiv:1512.06804 [hep-lat]} \BibitemShut {NoStop}%
\bibitem [{\citenamefont {Laio}\ \emph {et~al.}(2016)\citenamefont {Laio},
  \citenamefont {Martinelli},\ and\ \citenamefont {Sanfilippo}}]{Laio:2015era}%
  \BibitemOpen
  \bibfield  {author} {\bibinfo {author} {\bibfnamefont {A.}~\bibnamefont
  {Laio}}, \bibinfo {author} {\bibfnamefont {G.}~\bibnamefont {Martinelli}}, \
  and\ \bibinfo {author} {\bibfnamefont {F.}~\bibnamefont {Sanfilippo}},\
  }\href {\doibase 10.1007/JHEP07(2016)089} {\bibfield  {journal} {\bibinfo
  {journal} {JHEP}\ }\textbf {\bibinfo {volume} {07}},\ \bibinfo {pages} {089}
  (\bibinfo {year} {2016})},\ \Eprint {http://arxiv.org/abs/1508.07270}
  {arXiv:1508.07270 [hep-lat]} \BibitemShut {NoStop}%
\bibitem [{\citenamefont {Hasenbusch}(2017)}]{Hasenbusch:2017unr}%
  \BibitemOpen
  \bibfield  {author} {\bibinfo {author} {\bibfnamefont {M.}~\bibnamefont
  {Hasenbusch}},\ }\href {\doibase 10.1103/PhysRevD.96.054504} {\bibfield
  {journal} {\bibinfo  {journal} {Phys. Rev.}\ }\textbf {\bibinfo {volume}
  {D96}},\ \bibinfo {pages} {054504} (\bibinfo {year} {2017})},\ \Eprint
  {http://arxiv.org/abs/1706.04443} {arXiv:1706.04443 [hep-lat]} \BibitemShut
  {NoStop}%
\bibitem [{\citenamefont {Bonati}\ and\ \citenamefont
  {D'Elia}(2017)}]{Bonati:2017woi}%
  \BibitemOpen
  \bibfield  {author} {\bibinfo {author} {\bibfnamefont {C.}~\bibnamefont
  {Bonati}}\ and\ \bibinfo {author} {\bibfnamefont {M.}~\bibnamefont
  {D'Elia}},\ }\href@noop {} {\  (\bibinfo {year} {2017})},\ \Eprint
  {http://arxiv.org/abs/1709.10034} {arXiv:1709.10034 [hep-lat]} \BibitemShut
  {NoStop}%
\bibitem [{\citenamefont {{Cossu, Guido}}\ \emph {et~al.}(2018)\citenamefont
  {{Cossu, Guido}}, \citenamefont {{Boyle, Peter}}, \citenamefont {{Christ,
  Norman}}, \citenamefont {{Jung, Chulwoo}}, \citenamefont {{Jüttner,
  Andreas}},\ and\ \citenamefont {{Sanfilippo, Francesco}}}]{Cossu:2017eys}%
  \BibitemOpen
  \bibfield  {author} {\bibinfo {author} {\bibnamefont {{Cossu, Guido}}},
  \bibinfo {author} {\bibnamefont {{Boyle, Peter}}}, \bibinfo {author}
  {\bibnamefont {{Christ, Norman}}}, \bibinfo {author} {\bibnamefont {{Jung,
  Chulwoo}}}, \bibinfo {author} {\bibnamefont {{Jüttner, Andreas}}}, \ and\
  \bibinfo {author} {\bibnamefont {{Sanfilippo, Francesco}}},\ }\href {\doibase
  10.1051/epjconf/201817502008} {\bibfield  {journal} {\bibinfo  {journal} {EPJ
  Web Conf.}\ }\textbf {\bibinfo {volume} {175}},\ \bibinfo {pages} {02008}
  (\bibinfo {year} {2018})}\BibitemShut {NoStop}%
\bibitem [{\citenamefont {{Tu, Jiqun}}\ and\ \citenamefont {{Mawhinney,
  Robert}}(2018)}]{TuLattice2017}%
  \BibitemOpen
  \bibfield  {author} {\bibinfo {author} {\bibnamefont {{Tu, Jiqun}}}\ and\
  \bibinfo {author} {\bibnamefont {{Mawhinney, Robert}}},\ }\href {\doibase
  10.1051/epjconf/201817502006} {\bibfield  {journal} {\bibinfo  {journal} {EPJ
  Web Conf.}\ }\textbf {\bibinfo {volume} {175}},\ \bibinfo {pages} {02006}
  (\bibinfo {year} {2018})}\BibitemShut {NoStop}%
\bibitem [{\citenamefont {Endres}(2016)}]{Endres:2016rzj}%
  \BibitemOpen
  \bibfield  {author} {\bibinfo {author} {\bibfnamefont {M.~G.}\ \bibnamefont
  {Endres}},\ }\bibfield  {booktitle} {\emph {\bibinfo {booktitle}
  {{Proceedings, 34th International Symposium on Lattice Field Theory (Lattice
  2016): Southampton, UK, July 24-30, 2016}}},\ }\href@noop {} {\bibfield
  {journal} {\bibinfo  {journal} {PoS}\ }\textbf {\bibinfo {volume}
  {LATTICE2016}},\ \bibinfo {pages} {014} (\bibinfo {year} {2016})},\ \Eprint
  {http://arxiv.org/abs/1612.01609} {arXiv:1612.01609 [hep-lat]} \BibitemShut
  {NoStop}%
\bibitem [{\citenamefont {Borsanyi}\ \emph {et~al.}(2012)\citenamefont
  {Borsanyi} \emph {et~al.}}]{Borsanyi:2012zs}%
  \BibitemOpen
  \bibfield  {author} {\bibinfo {author} {\bibfnamefont {S.}~\bibnamefont
  {Borsanyi}} \emph {et~al.},\ }\href {\doibase 10.1007/JHEP09(2012)010}
  {\bibfield  {journal} {\bibinfo  {journal} {JHEP}\ }\textbf {\bibinfo
  {volume} {09}},\ \bibinfo {pages} {010} (\bibinfo {year} {2012})},\ \Eprint
  {http://arxiv.org/abs/1203.4469} {arXiv:1203.4469 [hep-lat]} \BibitemShut
  {NoStop}%
\bibitem [{\citenamefont {Cabibbo}\ and\ \citenamefont
  {Marinari}(1982)}]{CABIBBO1982387}%
  \BibitemOpen
  \bibfield  {author} {\bibinfo {author} {\bibfnamefont {N.}~\bibnamefont
  {Cabibbo}}\ and\ \bibinfo {author} {\bibfnamefont {E.}~\bibnamefont
  {Marinari}},\ }\href {\doibase
  http://dx.doi.org/10.1016/0370-2693(82)90696-7} {\bibfield  {journal}
  {\bibinfo  {journal} {Physics Letters B}\ }\textbf {\bibinfo {volume}
  {119}},\ \bibinfo {pages} {387 } (\bibinfo {year} {1982})}\BibitemShut
  {NoStop}%
\bibitem [{\citenamefont {Brown}\ and\ \citenamefont
  {Woch}(1987)}]{PhysRevLett.58.2394}%
  \BibitemOpen
  \bibfield  {author} {\bibinfo {author} {\bibfnamefont {F.~R.}\ \bibnamefont
  {Brown}}\ and\ \bibinfo {author} {\bibfnamefont {T.~J.}\ \bibnamefont
  {Woch}},\ }\href {\doibase 10.1103/PhysRevLett.58.2394} {\bibfield  {journal}
  {\bibinfo  {journal} {Phys. Rev. Lett.}\ }\textbf {\bibinfo {volume} {58}},\
  \bibinfo {pages} {2394} (\bibinfo {year} {1987})}\BibitemShut {NoStop}%
\bibitem [{\citenamefont {L{\"u}scher}(1982)}]{Luscher:1981zq}%
  \BibitemOpen
  \bibfield  {author} {\bibinfo {author} {\bibfnamefont {M.}~\bibnamefont
  {L{\"u}scher}},\ }\href {\doibase 10.1007/BF02029132} {\bibfield  {journal}
  {\bibinfo  {journal} {Commun. Math. Phys.}\ }\textbf {\bibinfo {volume}
  {85}},\ \bibinfo {pages} {39} (\bibinfo {year} {1982})}\BibitemShut {NoStop}%
\bibitem [{\citenamefont {Phillips}\ and\ \citenamefont
  {Stone}(1986)}]{Phillips:1986qd}%
  \BibitemOpen
  \bibfield  {author} {\bibinfo {author} {\bibfnamefont {A.}~\bibnamefont
  {Phillips}}\ and\ \bibinfo {author} {\bibfnamefont {D.}~\bibnamefont
  {Stone}},\ }\href {\doibase 10.1007/BF01211167} {\bibfield  {journal}
  {\bibinfo  {journal} {Commun. Math. Phys.}\ }\textbf {\bibinfo {volume}
  {103}},\ \bibinfo {pages} {599} (\bibinfo {year} {1986})}\BibitemShut
  {NoStop}%
\bibitem [{\citenamefont {'t~Hooft}(1995)}]{'tHooft1995491}%
  \BibitemOpen
  \bibfield  {author} {\bibinfo {author} {\bibfnamefont {G.}~\bibnamefont
  {'t~Hooft}},\ }\href {\doibase
  http://dx.doi.org/10.1016/0370-2693(95)00251-F} {\bibfield  {journal}
  {\bibinfo  {journal} {Physics Letters B}\ }\textbf {\bibinfo {volume}
  {349}},\ \bibinfo {pages} {491 } (\bibinfo {year} {1995})}\BibitemShut
  {NoStop}%
\bibitem [{\citenamefont {Asakawa}\ \emph {et~al.}(2015)\citenamefont
  {Asakawa}, \citenamefont {Hatsuda}, \citenamefont {Iritani}, \citenamefont
  {Itou}, \citenamefont {Kitazawa},\ and\ \citenamefont
  {Suzuki}}]{Asakawa:2015vta}%
  \BibitemOpen
  \bibfield  {author} {\bibinfo {author} {\bibfnamefont {M.}~\bibnamefont
  {Asakawa}}, \bibinfo {author} {\bibfnamefont {T.}~\bibnamefont {Hatsuda}},
  \bibinfo {author} {\bibfnamefont {T.}~\bibnamefont {Iritani}}, \bibinfo
  {author} {\bibfnamefont {E.}~\bibnamefont {Itou}}, \bibinfo {author}
  {\bibfnamefont {M.}~\bibnamefont {Kitazawa}}, \ and\ \bibinfo {author}
  {\bibfnamefont {H.}~\bibnamefont {Suzuki}},\ }\href@noop {} {\  (\bibinfo
  {year} {2015})},\ \Eprint {http://arxiv.org/abs/1503.06516} {arXiv:1503.06516
  [hep-lat]} \BibitemShut {NoStop}%
\bibitem [{\citenamefont {Edwards}\ and\ \citenamefont
  {Joo}(2005)}]{Edwards:2004sx}%
  \BibitemOpen
  \bibfield  {author} {\bibinfo {author} {\bibfnamefont {R.~G.}\ \bibnamefont
  {Edwards}}\ and\ \bibinfo {author} {\bibfnamefont {B.}~\bibnamefont {Joo}}
  (\bibinfo {collaboration} {SciDAC, LHPC, UKQCD}),\ }\bibfield  {booktitle}
  {\emph {\bibinfo {booktitle} {{Lattice field theory. Proceedings, 22nd
  International Symposium, Lattice 2004, Batavia, USA, June 21-26, 2004}}},\
  }\href {\doibase 10.1016/j.nuclphysbps.2004.11.254} {\bibfield  {journal}
  {\bibinfo  {journal} {Nucl. Phys. Proc. Suppl.}\ }\textbf {\bibinfo {volume}
  {140}},\ \bibinfo {pages} {832} (\bibinfo {year} {2005})},\ \Eprint
  {http://arxiv.org/abs/hep-lat/0409003} {arXiv:hep-lat/0409003 [hep-lat]}
  \BibitemShut {NoStop}%
\end{thebibliography}%
	
\end{document}